\documentclass{appolb}
\usepackage[utf8]{inputenc}
\usepackage{float}
\usepackage{sidecap}
\usepackage{bm}
\usepackage{wrapfig}
\usepackage{graphicx}
\usepackage{multirow}
\usepackage[T1]{fontenc}
\usepackage{dsfont}               
\usepackage{mathrsfs}             
\usepackage{slashed}              
\usepackage{amsmath}
\usepackage{amssymb}
\usepackage{amsbsy,nicefrac}
\usepackage{amsfonts}
\usepackage[usenames,dvipsnames]{color}

\usepackage{orcidlink}
\usepackage{graphicx,color,rotating}
\usepackage{dcolumn}
\usepackage{bm}
\newcommand{\er}{$\pm$}
\newcommand{\bc}           {\begin{center}}
\newcommand{\ec}           {\end{center}}
\newcommand{\bq}           {\begin{eqnarray}}
\newcommand{\eq}           {\end{eqnarray}}
\newcommand{\be}          {\begin{equation}}
\newcommand{\ee}           {\end{equation}}
\newcommand{\bi}           {\begin{itemize}}
\newcommand{\ei}           {\end{itemize}}
\definecolor{darkred}{rgb}{0.7,0,0}
\definecolor{darkgreen}{rgb}{0,0.4,0}
\definecolor{darkblue}{rgb}{0,0,0.4}
\definecolor{darkbrown}{rgb}{0.5,0,0}
\definecolor{darkcyan}{cmyk}{1,0.3,0.3,0.3}
\definecolor{midgreen}{rgb}{0,0.6,0}
\newcommand {\rd}{\color{red}}
\newcommand {\bl}{\color{blue}}
\newcommand {\gr}{\color{midgreen}}
\newcommand {\br}{\color{darkbrown}}

\usepackage{natbib}
\begin{document}

\title{\bf The Roper resonance and kin}%

\author{Volker Burkert
\address{Thomas Jefferson National Accelerator Facility, Newport News, VA, USA}
\\[3mm]
Eberhard Klempt 
\address{Helmholtz-Institute f\"ur Strahlen- und Kernphysik
der Rheinischen Friedrich-Wilhelms Universit\"at, Nussallee
14 - 16, 53115 Bonn, Germany\\
and\\
Thomas Jefferson National Accelerator Facility, Newport News, VA, USA} 
\\[5mm]}

\maketitle
\begin{abstract}

The properties of the Roper resonance $N(1440)\nicefrac{1}{2}^+$ are reviewed. Quark models have long struggled to reproduce its mass relative to its negative-parity partner $N(1535)\nicefrac{1}{2}^-$. This discrepancy motivated interpretations of the Roper as a dynamically generated meson–baryon state. Including its isospin partners $\Delta(1600)\nicefrac{3}{2}^+$ and $\Delta(1700)\nicefrac{3}{2}^-$ further accentuates the tension between quark-model predictions and experiment. Recent developments based on AdS/QCD and functional methods achieve much improved agreement, identifying the Roper as an ordinary three-quark excitation. Electroproduction experiments at Jefferson Lab have now resolved this long-standing question, revealing the Roper as a $qqq$ core dressed by a substantial meson cloud. The Roper resonance belongs to a family of four $N^\ast$ states with $J^P=\nicefrac{1}{2}^+$; the highest-mass member, $N(2100)\nicefrac{1}{2}^+$, likely represents a Roper-like excitation in the fourth shell.
\end{abstract}

\section{Introduction}
The Roper resonance — what a name for a particle! It is the only known resonance named after its discoverer.
Why is it important? The resonance is frequently and extensively discussed in the literature, yet it is comparatively rarely cited: 189 articles include the name “Roper” in their title, but only 40 of them cite David’s original paper\footnote{As of October 16, 2025}.
“The Roper” seems to have become as established a term as “the proton.” In both cases, there appears to be no need to cite the discoverer. But what makes $N(1440)\nicefrac12^+$ so special? 

In 1964, when David published his celebrated paper~\cite{Roper:1964zza}, baryon spectroscopy was still in its infancy.
More than a decade earlier, Enrico Fermi had used a $\pi^+$ meson beam scattering off protons in a hydrogen target~\cite{Anderson:1952nw}. The cross section showed a sharp rise from the pion-production threshold up to about 1200~MeV — a strong indication of the first baryon resonance, now known as $\Delta(1232)\nicefrac32^+$. The pion beam energy at that time, however, was insufficient to reveal the full resonance profile.

In the following years, a few other resonances were discovered. Each appeared as a distinct peak in the cross section and exhibited characteristic features in the observables that revealed their spin and parity. David’s resonance was different. He wrote:
“The resonance suggested in this paper, however, is not associated with conspicuous features in the observables measured so far and has been inferred from a more quantitative analysis.”
David’s work marked the beginning of a new era: the identification of resonances not through visible peaks, but through quantitative analyses. It became clear that resonances should not be defined merely by peaks in invariant mass spectra, but rather by studying the analytic behavior of complex scattering amplitudes.

The original motivation for David’s study of $\pi N$ interactions was to confirm a $P_{11}$ structure observed by Feld and Layson at a mass of about 1690\,MeV in the $\Lambda K$ invariant mass~\cite{Feld:1962dwa}. These authors found that the $\Lambda K$ peak was compatible with $\pi N$ scattering amplitudes. The $\Lambda K$ structure may well correspond to what we now identify as $N(1710)\nicefrac12^+$.

Later, David’s surprising result — that the $P_{11}$ wave of the $\pi N$ scattering amplitude resonates at a mass as low as 1440\,MeV — posed a serious challenge to the theoretical models that were subsequently developed. The 
low mass could not be easily explained within quark models and sparked a long-standing debate about its nature.

At the heart of this debate lies a fundamental question: What is a resonance?
Formally, we define resonances as poles in the complex scattering plane. But what physical mechanisms give rise to them? Are they generated by internal quark dynamics? Are all resonances of this kind? Or are some driven by molecular-type 
meson-baryon forces — perhaps all of them? Could every resonance be understood as a hadronic molecule?

\section{Properties of the Roper resonance} 
\begin{figure}[t!]
\vspace{-4mm}   
\hspace{-12mm}    
\includegraphics[width=1.2\linewidth]{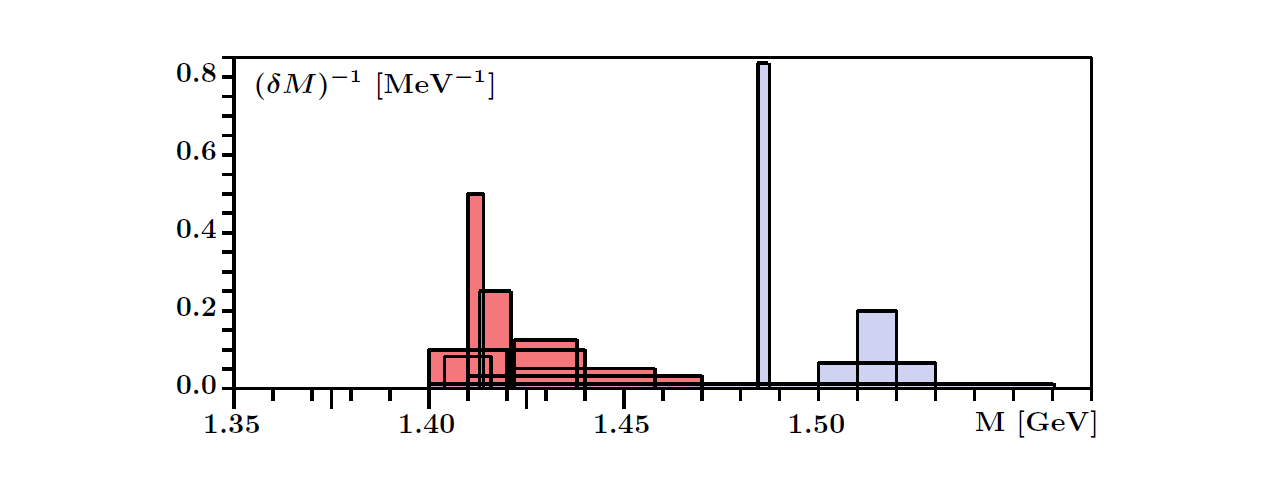}
\vspace{-8mm} 
\caption{Breit--Wigner masses from all entries in the RPP~\cite{ParticleDataGroup:2024cfk}. 
The box widths give the uncertainty, the height is inverse proportional to the uncertainty. 
The data are from 
Refs.~\cite{Arndt:2006bf,Batinic:2010zz,Cutkosky:1980rh,Hunt:2018wqz,Hohler:1979yr,Penner:2002ma,Sarantsev:2025lik,Shrestha:2012ep,%
CBELSATAPS:2015kka} (low mass)
and \cite{Arndt:2006bf,Shklyar:2012js,Vrana:1999nt} (high-mass).
See text for a discussion.}
    \label{fig:BW}
\end{figure}

The Roper resonance, or $N(1440)\nicefrac12^+$ as it is known today, has been studied in numerous experiments, often yielding partly inconsistent results. Figure~\ref{fig:BW} shows the distribution of measured masses as reported in the Review of Particle Physics (RPP)~\cite{ParticleDataGroup:2024cfk}. Each entry is represented by a box whose width corresponds to the quoted uncertainty and whose height is proportional to the inverse of that uncertainty.

The measurements cluster into two distinct groups: a high-mass cluster, primarily originating from data published before 2010, and a low-mass cluster, reported mostly after that year. Only Höhler~\cite{Hohler:1979yr,Hohler:1993lbk} and Cutkosky~\cite{Cutkosky:1980rh} had already reported low Roper masses at an early stage.

In determining the properties of the Roper resonance, results from publications reporting high Roper masses are excluded. However, all results listed in the RPP are included here, regardless of whether they were used in the averaging procedure.

Tables~\ref{tab:pole} and \ref{tab:BW} summarize the properties defined at the pole and in Breit–Wigner (BW) parameterizations: pole positions, helicity amplitudes at the pole, BW masses and widths, and BW helicity amplitudes.

\begin{table*}[t]
    \caption{Properties of the Roper resonance at the pole: Mass, widths,
    the elastic pole residue $r\exp{i\vartheta}$ ($r$ in MeV, $\vartheta$ in degree), and 
    helicity amplitudes (in $10^{-3}$\,GeV$^{-\nicefrac12}$). The last two lines
    gives the ranges suggested by the Particle Data Group and our mean values and their uncertainties.  \vspace{2mm} }
    \label{tab:pole}
    \footnotesize
    \centering
\renewcommand{\arraystretch}{1.3}
    \begin{tabular}{|cc|cc|cc|c|}
    \hline
$M_{\rm pole}$ & $\Gamma_{\rm pole}$&|r|& $\vartheta$&$A_{\nicefrac12} ^{\rm pole}$&phase&Ref.\\\hline
1366\er3   & 192\er4    &&& -60\er6  & -(30\er7)$^\circ$ & \cite{Sarantsev:2025lik}\\
1374\er5   & 215\er20   &58\er23&-(65\er11)$^\circ$&          &                   & \cite{Hoferichter:2023mgy}\\
1353\er1   & 203\er2    &59\er1&-(104\er2)$^\circ$& -90\er7  & -(30\er3)$^\circ$ & \cite{Ronchen:2022hqk}\\
1360       & 186        &49\er3&&          &                   & \cite{Hunt:2018wqz}\\
1370\er4   &  190\er7   &&& {\it 41\er5}   & {\it (23\er10)$^\circ$} & \cite{Anisovich:2017afs}\\
1369\er3   & 189\er5    &49\er3& -(82\er3)$^\circ$& -44\er5  & -(40\er8)$^\circ$ & \cite{CBELSATAPS:2015kka}\\
1355       & 215        & 62   & -98$^\circ$  & -60 & -23$^\circ$    &  \cite{Ronchen:2015vfa}\\
1363\er3   & 180\er6    &50\er2& -(88\er2)$^\circ$&          &                   & \cite{Svarc:2014zja}\\
1375\er30  & 180\er40   &52\er5& -(100\er35)$^\circ$&          &                   & \cite{Cutkosky:1980rh}\\
1385       &  164       & 40   &&          &                   & \cite{Hohler:1979yr,Hohler:1993lbk}\\\hline
1360-1380& 180-205       &50-60&&                &            & \cite{ParticleDataGroup:2024cfk}\\ 
1367\er 10 & 191\er16    &52\er7& -(90\er 15)$^\circ$& -65\er19 & -(33\er9)$^\circ$&  mean\\
\hline
\end{tabular}
\end{table*}
\begin{table*}[thb]
    \caption{Breit--Wigner properties of the Roper resonance: Masses, widths (in MeV), helicity amplitudes 
   $N(1440)\nicefrac12^+\to p\gamma$  (in $10^{-3}$\,GeV$^{-\nicefrac12}$), and branching ratios. The decay 
   into $N\rho$ can proceed with
    angular momentum $l=0$ and $\rho$ and nucleon spins add to $S=1/2$, or with $l=2, S=3/2$.     
    Helicity amplitudes for $N(1440)\nicefrac12^+\to n\gamma$ are given in italic.  \vspace{2mm} }
    \label{tab:BW}
    \footnotesize
    \centering
\renewcommand{\arraystretch}{1.3}
    \begin{tabular}{|cc|c|ccccc|c|}
    \hline
$M_{\rm BW}$&$\Gamma_{\rm BW}$&$A_{\nicefrac12} ^{\rm BW}$&$N\pi$&$\Delta\pi$&$N\rho_0$&$N\rho_2$&$Nf_0(500)$&Ref.\\\hline
1410\er10  & 290\er30   &-76\er8 &&15\er7&9\er4&9\er4&15\er5&\cite{Sarantsev:2025lik}\\
1417\er4   & 257\er11   &-91\er7 &59\er2&22\er4&1.3\er0.4&&16\er3&\cite{Hunt:2018wqz}\\
1430\er10  & 360\er30   &-61\er6 &63\er2&20\er7&&&15\er6& \cite{CBELSATAPS:2015kka}\\
1412\er2   &  248\er5   &-84\er3 & 85\er1 & 7\er1&1.3\er0.4 &&27\er1 &             \cite{Shrestha:2012ep}\\
1439\er19  & 437\er151  &        &62\er4 &&&&&\cite{Batinic:2010zz}\\
1440\er30  & 340\er70   &   &68\er4&&&&&\cite{Cutkosky:1980rh}\\
1410\er12  & 135\er10   & &51\er5&&&&&\cite{Hohler:1979yr}\\\hline
1410-1470   & 250-450     &-80 to -50 & 55-75 &&&&&\cite{ParticleDataGroup:2024cfk}\\ 
 1423\er13 & 295\er 96 & -78\er13 & 65\er10&16\er7&&&18\er7&mean\\
\hline
\end{tabular}|
\phantom{zz}\\
\begin{tabular}{|c|c|c|c|c|c|c|}
\hline
$A^{\rm BW} _{\nicefrac12\ N(1440)\to p\gamma}$\phantom{zz}\hspace{-2mm}&\hspace{-2mm}{\it 13\er12} \cite{Hunt:2018wqz}\hspace{-2mm}&\hspace{-2mm}{\it 53\er7} \cite{Anisovich:2017afs}%
\hspace{-2mm}&\hspace{-2mm}{\it 48\er4} \cite{Chen:2012yv}\hspace{-2mm}&\hspace{-2mm}%
{\it 35 to 55} \cite{ParticleDataGroup:2024cfk}\hspace{-1mm}&{\it 38\er22}&mean\\
\hline
\end{tabular}
\end{table*}
he reported values, e.g. for the pole mass, are statistically inconsistent. Most publications quote only statistical uncertainties, while systematic shifts in the mass can arise from various sources: amplitudes that do not fully satisfy theoretical constraints such as unitarity, analyticity, or crossing symmetry; incomplete inclusion or insufficient experimental constraints on allowed decay modes; or limitations in the model space, such as the number of resonances included across partial waves. These factors can all have a significant impact on the extracted results. We therefore estimate the overall uncertainty from the statistical spread of the published values.

The penultimate line lists the range of values defined by the PDG, while the final line gives the mean value and its associated uncertainty.

Most observables are consistently described with reasonably small uncertainties, with the notable exception of the $N(1440)\nicefrac12^+\to N\rho$ branching ratio. This quantity depends critically on the precise definition of the ratio~\cite{Burkert:2022bqo}. The partial decay width for the $S$-wave decay is given by
\begin{equation}
\Gamma_0(s) = \frac{g_0^2}{\sqrt{s_0}}\,\rho_0(s)\,\left(\frac{p}{p_0}\right)\,. \label{rel}
\end{equation}
For the $D$-wave decay, this term is multiplied by a centrifugal barrier factor. Conventionally, the branching ratio is defined at the nominal mass of the resonance, with
BR$_0=\Gamma_0(s_0)/\Gamma_{\rm tot}$,
and the partial width $\Gamma_0(s_0)=0$ due to the vanishing phase space.

However, the high-mass tail of the $N(1440)\nicefrac12^+$ resonance can decay into $N\rho$, particularly into 
the low-mass tail of $\rho$ mesons. This effect can be properly accounted for by integrating Eq.~(\ref{rel}) over the mass distributions of both the parent and daughter resonances.

\section{Interpretations of the Roper resonance}
The Roper resonance is the lowest-mass state with the quantum numbers of the nucleon. Only a few years after its discovery, it had already become the subject of considerable controversy. In Ref.~\cite{Argyres:1967zz}, the Roper resonance was interpreted as primarily an $N f_0(500)$\footnote{Referred to as S$_0$ with a mass of 700\,MeV} self-consistent bound state, consistent with the bootstrap picture of hadron resonances. In Ref.~\cite{Chylek:1968pn}, $N(1440)\nicefrac12^+$ was instead proposed to be the first radial excitation of the nucleon.

Since then, these two competing interpretations—whether the Roper resonance is a dynamically generated state or an ordinary three-quark excitation—have accompanied its study and shaped the ongoing discussion of its nature~\cite{Richard:2025nvn}.

\begin{figure}[pt]
    \label{tab:QM}
    \centering
\setlength{\unitlength}{0.45mm}
\linethickness{0.3mm}
\vspace{-20mm}
\begin{picture}(230.00,180.00)
\put(00.00,00.00){\line(1,0){240.00}}
\put(00.00,65.00){\line(1,0){240.00}}
\put(00.00,00){\line(0,1){65}} 
\put(240.00,00){\line(0,1){65}} 

\multiput(0,-3)(20,0){12}{\line(0,1){3}} 
\multiput(-3,0)(0,5){13}{\line(1,0){3}} 
\put(-20.00,65.00){\makebox(10.00,2.00)[l]{\bf\boldmath GeV}}
\put(-17.00,49.00){\makebox(10.00,2.00)[l]{\bf 1.6}}
\put(-17.00,29.00){\makebox(10.00,2.00)[l]{\bf 1.5}}
\put(-17.00,9.00){\makebox(10.00,2.00)[l]{\bf 1.4}}

\put(120.00,51.00){\makebox(10.00,2.00)[l]{\boldmath\br\footnotesize  $N(1440)$: $\ast$}}
\put(05.00,16.00){\makebox(10.00,2.00)[l]{\boldmath\tiny\br $N(1440)$}} 
\put(182.00,2.70){\makebox(10.00,2.00)[l]{\boldmath\tiny\br $N(1440)$}} 
\put(03.00,12.00){\br\line(1,0){177.00}}   
\put(03.00,24.00){\br\line(1,0){177.00}}   
\put(03.00,12.00){\br\line(0,1){12.00}}   
\put(179.00,12.00){\br\line(0,1){12.00}}   

\put(181.00,2.00){\br\line(1,0){59.00}}    
\put(181.00,6.00){\br\line(1,0){59.00}}    
\put(181.00,2.00){\br\line(0,1){4.00}}
\put(239.00,2.00){\br\line(0,1){4.00}}

\put(120.00,58.00){\makebox(10.00,2.00)[l]{\boldmath\rd\footnotesize  $N(1535)$: $\nabla$}}
\put(05.00,35.00){\makebox(10.00,2.00)[l]{\boldmath\tiny\rd $N(1535)$}}
\put(182.00,30.50){\makebox(10.00,2.00)[l]{\boldmath\tiny\rd $N(1535)$}}
\put(04.00,33.00){\rd\line(1,0){176.00}}  
\put(04.00,39.50){\rd\line(1,0){176.00}}  
\put(04.00,33.00){\rd\line(0,1){6.00}}
\put(179.00,33.00){\rd\line(0,1){6.00}}

\put(181.00,30.00){\rd\line(1,0){59.00}}  
\put(181.00,34.00){\rd\line(1,0){59.00}}  
\put(181.00,30.00){\rd\line(0,1){4.00}}
\put(239.00,30.00){\rd\line(0,1){4.00}}

\put(10.00,11.00){\makebox(10.00,2.00)[l]{\boldmath\br $\ast$}}    
\put(10.00,28.00){\makebox(10.00,2.00)[l]{\boldmath\rd $\nabla$}}  

\put(30.00,37.00){\makebox(10.00,2.00)[l]{\boldmath\br $\ast$}}  
\put(30.00,21.00){\makebox(10.00,2.00)[l]{\boldmath\rd $\nabla$}}  

\put(50.00,21.40){\makebox(10.00,2.00)[l]{\boldmath\br $\ast$}}  
\put(50.00,33.40){\makebox(10.00,2.00)[l]{\boldmath\rd $\nabla$}}  

\put(70.00,32.60){\makebox(10.00,2.00)[l]{\boldmath\br $\ast$}}  
\put(70.00,16.00){\makebox(10.00,2.00)[l]{\boldmath\rd $\nabla$}}  

\put(90.00,17.00){\makebox(10.00,2.00)[l]{\boldmath\br $\ast$}}    
\put(90.00,25.80){\makebox(10.00,2.00)[l]{\boldmath\rd $\nabla$}}  

\put(110.00,21.70){\makebox(10.00,2.00)[l]{\boldmath\br $\ast$}}    
\put(110.00,34.00){\makebox(10.00,2.00)[l]{\boldmath\rd $\nabla$}}  

\put(129.00,37.70){\makebox(10.00,2.00)[l]{\boldmath\br $\ast$}}    
\put(131.00,36.70){\makebox(10.00,2.00)[l]{\boldmath\rd $\nabla$}}  

\put(150.00,31.20){\makebox(10.00,2.00)[l]{\boldmath\br $\ast$}}    
\put(150.00,36.70){\makebox(10.00,2.00)[l]{\boldmath\rd $\nabla$}}  

\put(170.00,17.00){\makebox(10.00,2.00)[l]{\boldmath\br $\ast$}}    
\put(170.00,42.20){\makebox(10.00,2.00)[l]{\boldmath\rd $\nabla$}}  

\put(190.00,6.20){\makebox(10.00,2.00)[l]{\boldmath\br $\ast$}}     
\put(190.00,36.00){\makebox(10.00,2.00)[l]{\boldmath\rd $\nabla$}}  

\put(210.00,3.40){\makebox(10.00,2.00)[l]{\boldmath\br $\ast$}}     
\put(212.00,43.40){\makebox(10.00,2.00)[l]{\boldmath\rd $\nabla$}}  

\put(230.00,24.00){\makebox(10.00,2.00)[l]{\boldmath\br $\ast$}}    
\put(232.00,42.00){\makebox(10.00,2.00)[l]{\boldmath\rd $\nabla$}}  
\end{picture}
\begin{picture}(230.00,115.00)
\put(00.00,20.00){\line(1,0){240.00}}
\put(00.00,105.00){\line(1,0){240.00}}
\put(00.00,20){\line(0,1){85}} 
\put(240.00,20){\line(0,1){85}} 

\multiput(0,17)(20,0){12}{\line(0,1){3}} 
\multiput(-3,20)(0,5){17}{\line(1,0){3}} 
\put(-20.00,105.00){\makebox(10.00,2.00)[l]{\bf\boldmath GeV}}
\put(-17.00,89.00){\makebox(10.00,2.00)[l]{\bf 1.8}}
\put(-17.00,69.00){\makebox(10.00,2.00)[l]{\bf 1.7}}
\put(-17.00,49.00){\makebox(10.00,2.00)[l]{\bf 1.6}}
\put(-17.00,29.00){\makebox(10.00,2.00)[l]{\bf 1.5}}

\put(120.00,91.00){\makebox(10.00,2.00)[l]{\boldmath\gr\footnotesize  $\Delta(1600)$: $\Delta$}}
\put(05.00,44.00){\makebox(10.00,2.00)[l]{\boldmath\tiny\gr $\Delta(1600)$}}
\put(182.00,44.00){\makebox(10.00,2.00)[l]{\boldmath\tiny\gr $\Delta(1600)$}}
\put(03.00,30.00){\gr\line(1,0){177.00}}  
\put(03.00,58.00){\gr\line(1,0){177.00}}  
\put(03.00,30.00){\gr\line(0,1){28.00}}
\put(180.00,30.00){\gr\line(0,1){28.00}}

\put(180.00,24.00){\gr\line(1,0){59.00}}  
\put(180.00,48.00){\gr\line(1,0){59.00}}  
\put(180.00,24.00){\gr\line(0,1){24.00}}
\put(239.00,24.00){\gr\line(0,1){24.00}}

\put(120.00,98.00){\makebox(10.00,2.00)[l]{\boldmath\bl\footnotesize  $\Delta(1700)$: $\otimes$}}
\put(05.00,71.00){\makebox(10.00,2.00)[l]{\boldmath\tiny\bl $\Delta(1700)$}}
\put(182.00,63.00){\makebox(10.00,2.00)[l]{\boldmath\tiny\bl $\Delta(1700)$}}
\put(03.00,68.00){\bl\line(1,0){177.00}}  
\put(03.00,76.00){\bl\line(1,0){177.00}}  
\put(03.00,68.00){\bl\line(0,1){8.00}}
\put(180.00,68.00){\bl\line(0,1){8.00}}

\put(180.00,58.00){\bl\line(1,0){59.00}}  
\put(180.00,68.00){\bl\line(1,0){59.00}}  
\put(180.00,58.00){\bl\line(0,1){10.00}}
\put(239.00,58.00){\bl\line(0,1){10.00}}

\put(10.00,86.00){\makebox(10.00,2.00)[l]{\boldmath\gr $\Delta$}}  
\put(10.00,67.00){\makebox(10.00,2.00)[l]{\boldmath\bl $\otimes$}} 
\put(-5.00,3.00){\makebox(10.00,2.00)[l]{\cite{Isgur:1978xj,Isgur:1978wd}}}
\put(5.00,12.00){\makebox(10.00,2.00)[l]{\footnotesize OGE}}

\put(30.00,88.00){\makebox(10.00,2.00)[l]{\boldmath\gr $\Delta$}}  
\put(30.00,53.00){\makebox(10.00,2.00)[l]{\boldmath\bl $\otimes$}}  
\put(25.00,3.00){\makebox(10.00,2.00)[l]{\cite{Capstick:1986ter}}}
\put(25.00,12.00){\makebox(10.00,2.00)[l]{\footnotesize OGE}}

\put(50.00,74.00){\makebox(10.00,2.00)[l]{\boldmath\gr $\Delta$}}  
\put(50.00,59.40){\makebox(10.00,2.00)[l]{\boldmath\bl $\otimes$}}  
\put(45.00,3.00){\makebox(10.00,2.00)[l]{\cite{Glozman:1997ag}}}
\put(45.00,12.00){\makebox(10.00,2.00)[l]{\footnotesize OBE}}

\put(70.00,102.20){\makebox(10.00,2.00)[l]{\boldmath\gr $\Delta$}} 
\put(70.00,54.60){\makebox(10.00,2.00)[l]{\boldmath\bl $\otimes$}}  
\put(65.00,3.00){\makebox(10.00,2.00)[l]{\cite{Loring:2001kx}}}
\put(65.00,12.00){\makebox(10.00,2.00)[l]{\footnotesize I.I.I.}}

\put(86.00,47.80){\makebox(10.00,2.00)[l]{\boldmath\gr $\Delta$}}  
\put(93.00,48.20){\makebox(10.00,2.00)[l]{\boldmath\bl $\otimes$}} 
\put(85.00,3.00){\makebox(10.00,2.00)[l]{\cite{Ronniger:2012xp}}}
\put(85.00,12.00){\makebox(10.00,2.00)[l]{\footnotesize I.I.I.}}

\put(110.00,74.40){\makebox(10.00,2.00)[l]{\boldmath\gr $\Delta$}}  
\put(110.00,43.60){\makebox(10.00,2.00)[l]{\boldmath\bl $\otimes$}} 
\put(105.00,3.00){\makebox(10.00,2.00)[l]{\cite{Giannini:2001kb}}}
\put(105.00,12.00){\makebox(10.00,2.00)[l]{\footnotesize HQM}}

\put(130.00,71.80){\makebox(10.00,2.00)[l]{\boldmath\gr $\Delta$}}  
\put(130.00,63.70){\makebox(10.00,2.00)[l]{\boldmath\bl $\otimes$}} 
\put(125.00,3.00){\makebox(10.00,2.00)[l]{\cite{Santopinto:2004hw}}}
\put(125.00,12.00){\makebox(10.00,2.00)[l]{\footnotesize DQ}}

\put(150.00,66.80){\makebox(10.00,2.00)[l]{\boldmath\gr $\Delta$}}  
\put(150.00,54.00){\makebox(10.00,2.00)[l]{\boldmath\bl $\otimes$}} 
\put(145.00,3.00){\makebox(10.00,2.00)[l]{\cite{Santopinto:2014opa}}}
\put(145.00,12.00){\makebox(10.00,2.00)[l]{\footnotesize DQ}}

\put(168.00,58.20){\makebox(10.00,2.00)[l]{\boldmath\gr $\Delta$}}  
\put(172.00,58.80){\makebox(10.00,2.00)[l]{\boldmath\bl $\otimes$}} 
\put(165.00,3.00){\makebox(10.00,2.00)[l]{\cite{Bijker:1994yr}}}
\put(165.00,12.00){\makebox(10.00,2.00)[l]{\footnotesize AM}}

\put(188.00,51.60){\makebox(10.00,2.00)[l]{\boldmath\gr $\Delta$}}  
\put(192.00,51.60){\makebox(10.00,2.00)[l]{\boldmath\bl $\otimes$}} 
\put(185.00,3.00){\makebox(10.00,2.00)[l]{\cite{Brodsky:2014yha}}}
\put(185.00,13.00){\makebox(10.00,2.00)[l]{\footnotesize AdS}}

\put(208.00,43.00){\makebox(10.00,2.00)[l]{\boldmath\gr $\Delta$}}  
\put(210.00,61.40){\makebox(10.00,2.00)[l]{\boldmath\bl $\otimes$}} 
\put(205.00,3.00){\makebox(10.00,2.00)[l]{\cite{Burkert:2025coj}}}
\put(205.00,12.00){\makebox(10.00,2.00)[l]{\footnotesize MF}}

\put(228.00,46.00){\makebox(10.00,2.00)[l]{\boldmath\gr $\Delta$}}  
\put(230.00,52.00){\makebox(10.00,2.00)[l]{\boldmath\bl $\otimes$}} 
\put(225.00,3.00){\makebox(10.00,2.00)[l]{\cite{Eichmann:2016hgl}}}
\put(225.00,12.00){\makebox(10.00,2.00)[l]{\footnotesize FM}}
\end{picture}
\vspace{2mm}
\caption{Comparison of calculated masses of spin-$\nicefrac12$ nucleon and spin-$\nicefrac32$ $\Delta$ resonances
    with RPP values (shown by boxes). The models are characterized by their interaction: One-gluon exchange (OGE)~\cite{Capstick:1986ter}, 
    Goldstone-boson exchange (GBE)~\cite{Glozman:1997ag},
    instanton-induced interactions (I.I.I.) without~\cite{Loring:2001kx} or with additional Goldstone-boson 
    exchange~\cite{Ronniger:2012xp}, the hypercentral constituent-quark model (HQM)~\cite{Giannini:2001kb},
    diquark (DQ) models~\cite{Santopinto:2004hw,Santopinto:2014opa}, the algebraic model (AM)~\cite{Bijker:1994yr},
    a holographic model~\cite{Brodsky:2014yha}, an empirical mass formula (MF)~\cite{Burkert:2025coj},
    and functional methods (FM)~\cite{Eichmann:2016hgl}
    The three latter models are compared to the pole masses. \vspace{2mm} \label{masses} }
\end{figure}
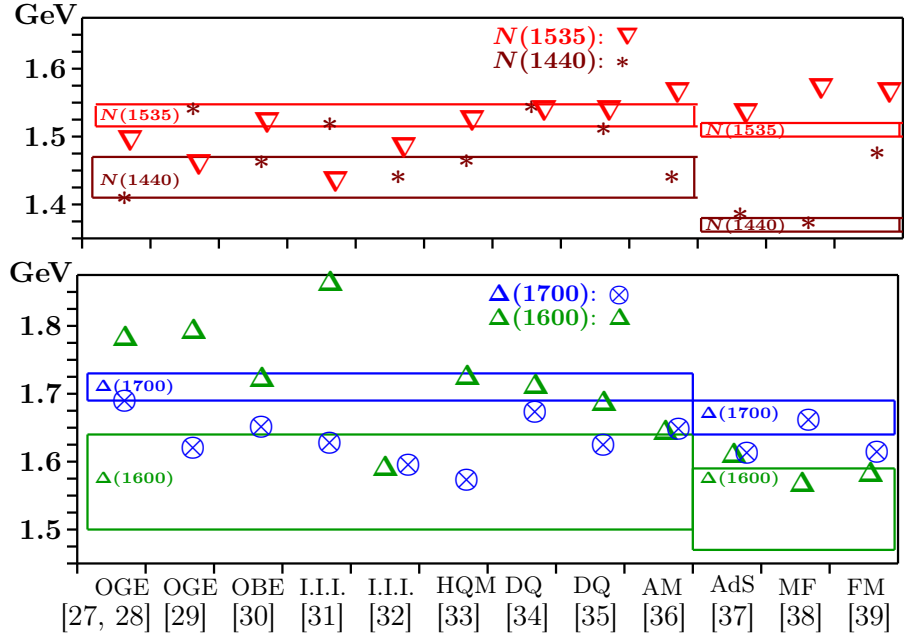 

In the first consistent quark-model calculation of the baryon resonance spectrum, Isgur and Karl assumed a linear confining potential, combined with an effective one-gluon exchange~\cite{Isgur:1978xj,Isgur:1978wd}. A large number of alternative models were subsequently developed. One serious problem soon emerged: as a three-quark state, the Roper resonance—belonging to the second excitation shell—appeared with a lower mass than $N(1535)\nicefrac12^-$, which belongs to the first excitation shell. This inversion is difficult to explain.

In Fig.~\ref{masses}, the experimental masses of the Roper resonance and its nearby negative-parity partner, $N(1535)\nicefrac12^-$, are compared with theoretical predictions. Model parameters are often tuned to reproduce the experimental values; therefore, we also compare the corresponding pair of resonances in the $\Delta$ sector: $\Delta(1600)\nicefrac32^+$, the first radial excitation of $\Delta(1232)\nicefrac32^+$, and $\Delta(1700)\nicefrac32^-$, the first orbital excitation with $J=3/2$.

In most dynamical models, the mass ordering $M_{N(1440)\nicefrac12^+} > M_{N(1535)\nicefrac12^-}$ is obtained, contrary to experiment. In those cases where the ordering is correct, $M_{\Delta(1600)\nicefrac32^+} > M_{\Delta(1700)\nicefrac32^-}$, the predicted $\Delta(1600)\nicefrac32^+$ mass is usually considerably too high. Reasonable agreement with the data is achieved only in Refs.~\cite{Bijker:1994yr,Burkert:2025coj} and \cite{Brodsky:2014yha,Eichmann:2016hgl}. The latter two are QCD-based approaches starting from quarks with (nearly) vanishing current masses.

These two QCD-based models are conceptually very different. Ref.~\cite{Brodsky:2014yha} employs light-front holographic QCD, exploiting the correspondence between gravity in anti–de Sitter (AdS) space and a conformal field theory. Ref.~\cite{Eichmann:2016hgl}, by contrast, uses functional methods derived from the path integral formalism, converting classical equations of motion (Klein–Gordon, Dirac, and Maxwell equations) into their quantum counterparts such as the Dyson–Schwinger equations. Both approaches are firmly grounded in QCD. Their results demonstrate that the low mass of the Roper resonance should no longer be regarded as an unresolved problem.

The difficulties of early models in reproducing the low masses of the Roper and $\Delta(1600)\nicefrac32^+$ resonances led to interpretations of $N(1440)\nicefrac12^+$ beyond the traditional quark model. In the bag model, the Roper may represent a collective vibration of the bag surface. Indeed, within the Skyrme model—where baryons appear as topological solitons of nonlinear meson fields—a resonance was found in the breathing mode at $M = 1420$\,MeV~\cite{Kaulfuss:1985na}. Alternatively, the Roper could have a large gluonic component~\cite{Li:1991yba}, denoted as $qqqg$. Such “hybrid” states were two decades later predicted by lattice QCD, although at significantly higher masses~\cite{Dudek:2012ag}.

The authors of Ref.~\cite{Krehl:1999km} investigated the Roper resonance within a coupled-channel meson-exchange model including $\Delta(1232)\pi$, $N\rho$, and $Nf_0(500)$ channels. They found that the Roper resonance can be described purely by meson–baryon dynamics—without the need to introduce an explicit three-quark core—to reproduce $\pi N$ phase shifts and inelasticities. The resonance thus emerges dynamically. These results were confirmed in Refs.~\cite{Ronchen:2012eg,Ronchen:2014cna}, where both $N(1440)\nicefrac12^+$ and $\Delta(1600)\nicefrac32^+$ were dynamically generated without requiring genuine three-quark components. Similarly, Ref.~\cite{Owa:2025mep} identifies the Roper as a dynamically generated state with negligible contributions from bare states, while Ref.~\cite{Hockley:2024ipz} finds that the dominant contributions to $\Delta(1600)\nicefrac32^+$ arise from strong rescattering in the $\pi N$ and $\pi\Delta(1232)$ channels, interpreting $\Delta(1600)\nicefrac32^+$ as a dynamically generated resonance.

The internal structure of the Roper resonance—whether it is a three-quark state, a hybrid, or a meson–baryon molecule—cannot be inferred from its mass alone. Its true nature can only be determined through electroproduction experiments spanning a wide range of momentum transfers.

\section{Electroproduction of the Roper resonance}

We first briefly discuss how meson electroproduction contributes to elucidating the nature of the excited states. The virtual (spacelike) photon exchanged between the scattered electron and the target nucleon has a finite lifetime and, correspondingly, probes the nucleon’s interior within a limited spatial domain. The higher the photon virtuality $Q^2$, the shorter its lifetime, and the finer the spatial resolution that can be achieved.

Excitation of the ground-state nucleon through electron scattering in the s-channel provides a powerful means of investigating the structural properties of the excited states. The Roper resonance, as the lowest-mass nucleon resonance, exhibits many features—such as structural complexity—comparable to those of higher-mass resonances, and may thus serve as a representative example of the latter.

In the following, we employ the $Q^2$-dependence of the excitation strength as a tool to disentangle the various contributions to the resonance strength and to identify the respective roles of the quark core and meson–baryon components.
\begin{figure}[!t!]
\centering
\includegraphics[width=0.7\linewidth]{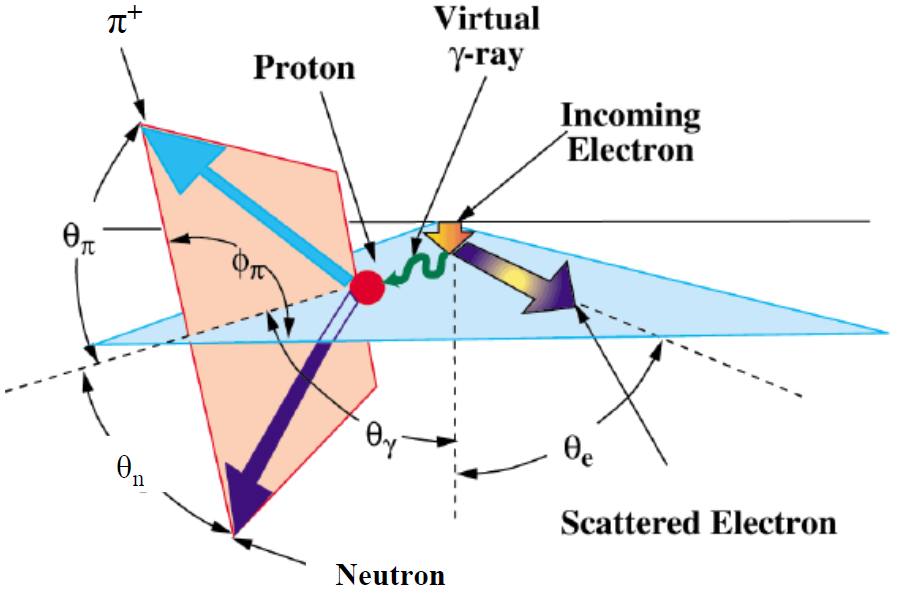}
\caption{Kinematics of $\pi^+$ electroproduction off protons in the laboratory system.}
\label{kine}
\end{figure}
The Roper resonance, as a state with spin-parity $J^P = \nicefrac{1}{2}^+$ and isospin $\nicefrac{1}{2}$, can be studied most effectively on a proton target in the reaction $e p \to e^\prime \pi^+ n$. The kinematics for single $\pi^+$ production are shown in Fig.~\ref{kine}.
\begin{figure}[!bt]
\centering
\hspace{-7mm}
\includegraphics[width=1.0\linewidth]{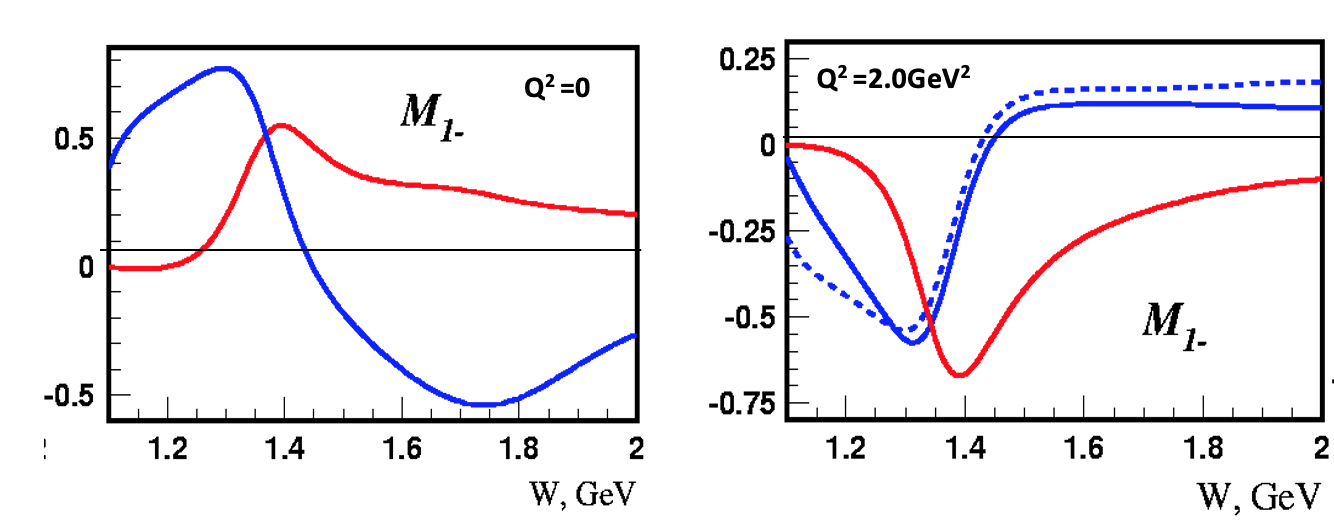}
    \caption{Magnetic transition multipole $M_{1-}$ of the proton to Roper resonance transition versus W. The red lines represent the imaginary part of  $M_{1-}$, the blue lines the real part, using two different analysis techniques; left: $Q^2=0$, right: $Q^2=2$\,GeV$^2$.} 
    \label{M1-}
\end{figure}
The unpolarized differential cross section has four terms, or five if the beam electrons are spin polarized. These terms relate to the transverse photon absorption cross section $\sigma_T$, the longitudinal photon absorption cross section $\sigma_L$, and the transverse-transverse interference terms $\sigma_{TT}$ and transverse-longitudinal term $\sigma_{TL}$. Measurement of all these cross sections in large acceptance detectors such as the CLAS at Jefferson Lab enables a detailed breakdown of the different electromagnetic contributions into the various partial wave elements underlying the production process. In the case of Roper-like resonances  the magnetic $M_{1-}$ and the scalar $S_{1-}$ are the relevant contributions leading to real and imaginary parts of the amplitude. The magnetic transition amplitude $M_{1-}$ is shown in Fig.~\ref{M1-} for two values of $Q^2=0$ and $Q^2=2$\,GeV$^2$, and as a function of the invariant mass $W=M(n\pi^+)$. 

It shows that the resonance character of the Roper $N(1440)\nicefrac{1}{2}^+$ is more prominently visible at  high $Q^2$ in both the real and  imaginary parts of the amplitude in comparison with the situation at the $Q^2=0$. 
The resonance position is clearly seen in the peak of the imaginary parts $M_{1-}$ at both values of $Q^2$. However, at $Q^2=0$ the real part clearly shows stronger and more extended contributions, which are not seen at the higher $Q^2$ value. This indicates the effects of more complex and extended contributions of meson-baryon both in the imaginary, but especially in the real part of the $M_{1-}$, while at high $Q^2$ the coupling is consistent with to a well-localized and spatially compact three-quark core. 
\begin{figure}[!th]
\centering
\hspace{-4mm}\includegraphics[width=1.0\linewidth]{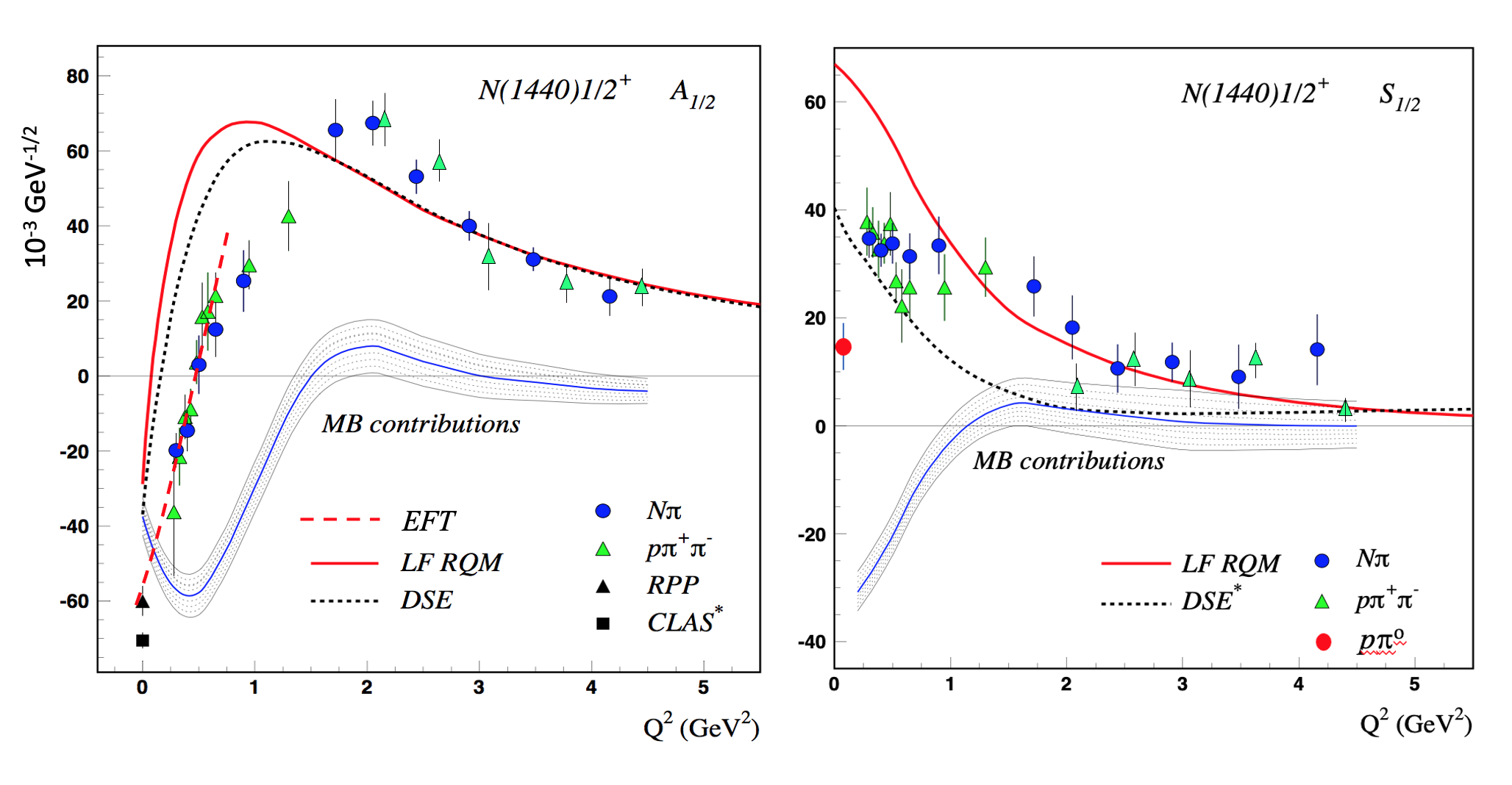}
\caption{\small Left: The $\gamma^\ast p\to N(1440)$ helicity transition amplitude $A_{1/2}$. Right: $S_{1/2}$. Both amplitudes in units of $10^{-3}$\,GeV$^{-1/2}$. 
The solid red curves are from a light-front relativistic quark model~\cite{Aznauryan:2012ec} and the black dashed curves are DSE results~\cite{Segovia:2015hra}.
 The green triangles and the blue full circles markers are data from the CLAS collaboration~\cite{CLAS:2012wxw,CLAS:2009ces,Mokeev:2023zhq}. The red cicle 
at the very small $Q^2$ value is from~\cite{Stajner:2017fmh}. See also text for details.}
\label{p11} 
\end{figure}
Although the partial-wave analysis clearly shows the sign change and the strong excitation of the Roper resonance at the real-photon point and at high $Q^2$, a separation of the resonant and non-resonant contributions is required in order to compare the data with theoretical predictions and advanced quark-based models that do not include meson–baryon effects. The procedure used to separate the resonant parts of the magnetic and scalar transition multipoles $M_{1-}$ and $S_{1-}$ and to relate them to the amplitudes $A_{1/2}$ and $S_{1/2}$ is described in detail in~\cite{CLAS:2009ces,Aznauryan:2011qj}.

In Fig.~\ref{p11}, $A_{1/2}$ and $S_{1/2}$ are displayed as functions of $Q^2$, where $A_{1/2}$ represents the transition strength of the transverse virtual photon and $S_{1/2}$ that of the scalar (or longitudinal) virtual photon. Applying the Siegert theorem~\cite{Siegert:1937yt} in the long-wavelength limit leads to the constraint $S_{1/2}=0$ at $\vec{Q}^2=0$, corresponding to the nonphysical region ($Q^2 < 0$). This partly explains the steep drop of the $S_{1/2}$ amplitude near the real-photon point.

The $Q^2$ dependence of the $A_{1/2}$ amplitude is particularly interesting. It starts at a large negative value near the real-photon point ($Q^2=0$) and rises steeply with increasing $Q^2$, changing sign near $Q^2 \approx 0.5$,GeV$^2$. A sign change in a transition amplitude (or transition form factor) is often associated with a node in the corresponding radial wave function and can therefore be interpreted as a first indication of the radial structure of the Roper resonance. We also note that this sign change in the $A_{1/2}$ amplitude is predicted by both theoretical approaches considered—the Dyson–Schwinger equation (DSE) framework and the light-front relativistic quark model (LF RQM). Both calculations are based on the underlying quark structure of the resonance and are expected to reproduce the region $Q^2 > 1$–$2$,GeV$^2$, while the low-$Q^2$ domain, where meson–baryon contributions are significant, can only be described qualitatively within such models.

This is indeed observed: in both theoretical approaches the node occurs at smaller values of $Q^2$ than in the data. This discrepancy is not unexpected, as the strong meson–baryon components in the Roper’s wave function at low $Q^2$, which are not included in these models, are likely to affect the dynamics of the resonance transition.

Following the sign change, the transverse amplitude reaches a maximum positive value near $Q^2 \approx 1.5$,GeV$^2$, before gradually decreasing at higher $Q^2$. In this kinematic region, good agreement is observed between both theoretical predictions and the data. The meson–baryon contributions diminish rapidly above $Q^2 > 2$\,GeV$^2$, as indicated by the shaded areas in Fig.~5.

For the scalar (longitudinal) transition amplitude $S_{1/2}$, a different and simpler trend is seen. Except for the very small $Q^2$ range discussed above, the amplitude decreases smoothly with increasing $Q^2$ and agrees with the LF RQM predictions for $Q^2 > 1.5$\,GeV$^2$, consistent with the behavior of $A_{1/2}$. In contrast, the DSE approach predicts a significantly steeper falloff with $Q^2$ than is observed in the data.

\begin{figure}[!t!]
\centering\includegraphics[width=0.7\linewidth]{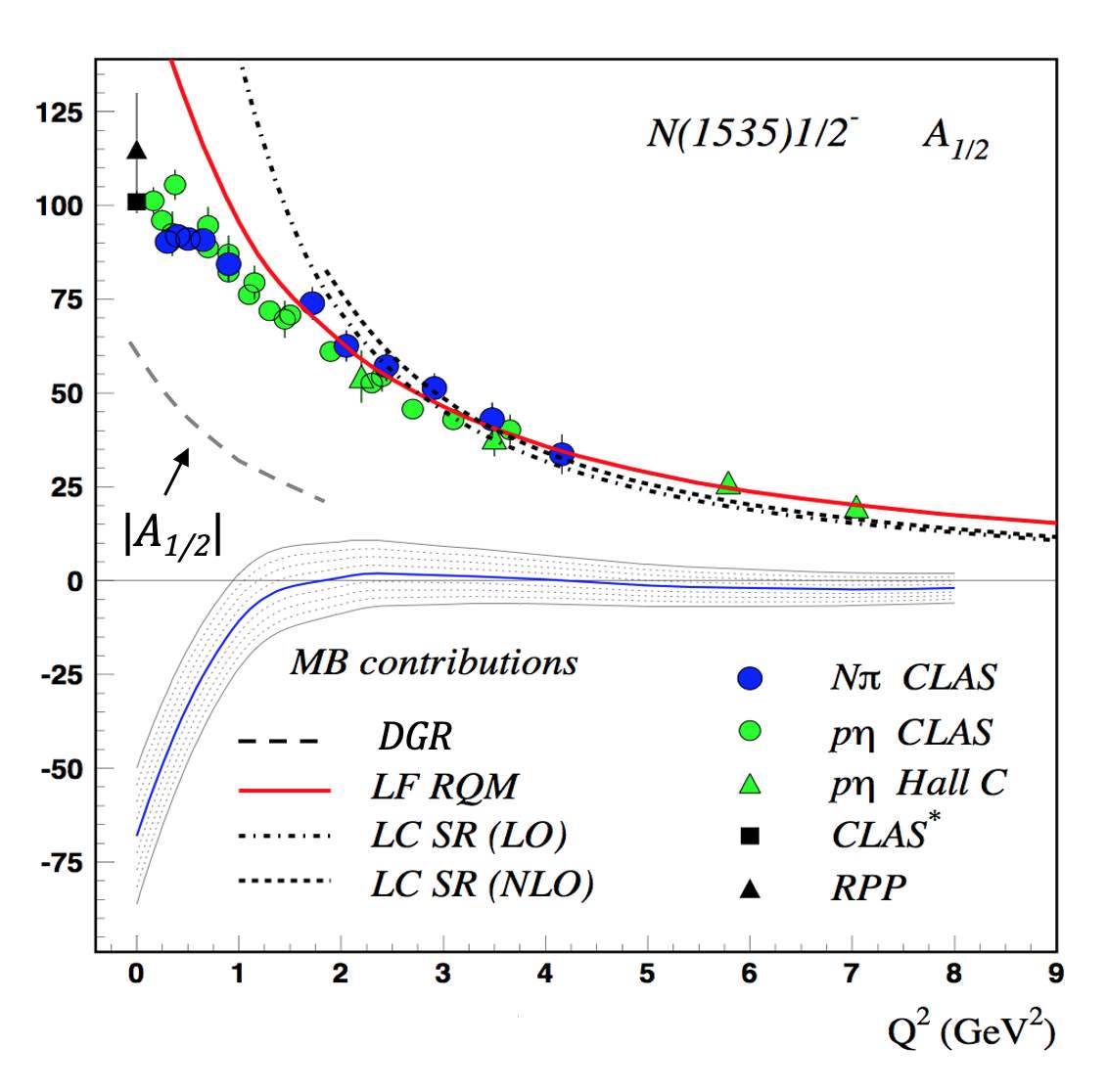}
\caption{\small Transition amplitude $A_{1/2}$ of $N(1535)\nicefrac{1}{2}^-$. The theory curves are from the {\it LF RQM}~\cite{Aznauryan:2012ec} (solid-red line), and the light-cone sum rule (dotted lines)~\cite{Anikin:2015ita} ({\it LC SR}) QCD approach. The dashed lines represent a dynamically generated resonance model~\cite{Jido:2007sm}.  }
\label{s11} 
\end{figure}
\subsection{Comparison of the Roper $N(1440)\nicefrac{1}{2}^+$ with $N(1535)\nicefrac{1}{2}^-$}

The $N(1535)\nicefrac{1}{2}^-$ resonance in the quark model is the first orbital excitation of the proton with one of the three quarks in an orbital state with $L_q=1$ and negative parity. In the quark model this state is thus the parity partner to the nucleon. Furthermore, similar to the Roper resonance it is also considered a dominantly dynamically generated resonance~\cite{Jido:2007sm} or as a meson-baryon molecule~\cite{Bruns:2010sv,Kaiser:1995cy}. Fortunately, there are data available at Jefferson Lab, both from CLAS in Hall B and from Hall C, that extend over the largest range in $Q^2$ available for any excited baryon resonance. These results are shown in Fig.~\ref{s11}. Both quark-based approaches show excellent agreement with the data at $Q^2 > 1-2~$\,GeV$^2$ and confirm the presence of a strong quark core as probed in electroproduction at the shorter distance scales. 
The attempt to describe the transition amplitude solely within a dynamically generated resonance models (DGR), underestimates the resonance strength even at the lowest $Q^2$ where dynamical processes may make significant contributions, and drops much more rapidly with $Q^2$ than the data, as expected for meson-baryon resonance contributions with a minimum of five quarks involved in the interaction compared to the minimum of the three quarks at the core of the quark model.   

Such data-based observations have contributed to today's prevailing view that the Roper resonance as well as  $N^{\star}$ and $\Delta^\star$ resonances in the mass range below $W \sim 1.8$\,GeV all contain a three-quark core that defines the characteristics of the resonance at all length scales, and  meson-baryon components involving five or more quarks. The latter component occupy distances close to the periphery of the excited state, while the former provides the quark core at distances close to the center.  
\begin{figure}[h!]
\centering\includegraphics[width=1.0\linewidth]{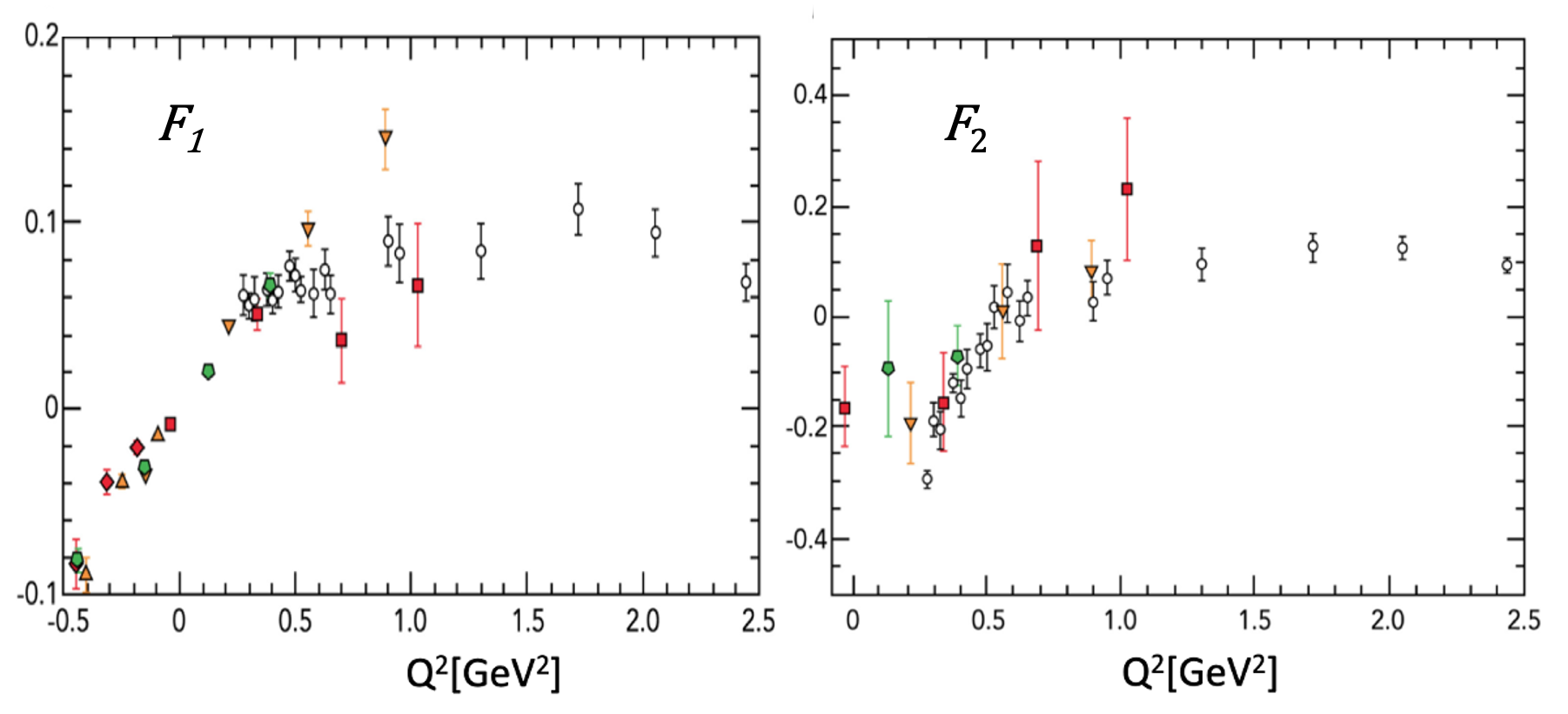}
\caption{\small Roper resonance transition form factor. Left $F_1(t)$, right: $F_2(t)$. The open circles are the CLAS data. The other (colored) symbols represent LQCD data. Lattice results are for pion masses 390~MeV (red squares), 450~MeV (orange triangles) and
875~MeV (green pentagons). }
\label{Roper-LQCD} 
\end{figure}
\subsection{Structure of the Roper resonance from Lattice QCD. }
\label{LQCD}
The structure of the $N(1440)\nicefrac{1}{2}^+$ resonance as represented with the transition form factors $F_1$ and $F_2$, has also been investigated in lattice QCD simulations, albeit at relatively large pion masses from 390 to 875 MeV~\cite{Lin:2011da}. The results are shown in Fig.~\ref{Roper-LQCD}.  In kinematics where the simulation overlap with the data, for the lowest pion mass one sees an agreement of the lattice calculations with the experimental results.   
\section{Higher-mass scalar excitations of the nucleon}

In quark models, baryons are described as systems of three constituent quarks, each with spin $\nicefrac12$.
The total baryonic wave function must be antisymmetric with respect to the exchange of any two quarks. The color wave function is completely antisymmetric ($A$), while the combined space–-spin–- flavor wave function is symmetric ($S$).

If the three quarks have identical masses, their spin–flavor wave functions obey SU(6) symmetry. The flavor wave functions of the nucleon and its excitations are characterized by isospin $I=\nicefrac12$ and exhibit mixed symmetry ($\tau_{M_S, M_A}$). The spin wave function $\chi$ can also have mixed symmetry ($\chi_{M_S, M_A}$ for spin $S=\nicefrac12$) or be symmetric ($\chi_{S}$ for $S=\nicefrac32$).

The spin–flavor wave functions can therefore assume the configurations $^28[56]$, $^28[70]$, $^48[70]$, and $^28[20]$, where the [56]-plet is fully symmetric, the [70]-plets have mixed symmetry, and the [20]-plet is fully antisymmetric.

The spatial dynamics of the three quarks are described by the time-dependent Jacobi coordinates
$\vec\rho=\frac{1}{\sqrt 2} (\vec r_1-\vec r_2)$,
$\vec\lambda=\frac{1}{\sqrt 6} (\vec r_1+\vec r_2-2\vec r_3)$,
and the trivial center-of-mass coordinate.
In the harmonic-oscillator (h.o.) approximation, both oscillators can be excited with orbital angular momenta $l_\rho$ and $l_\lambda$, and can also undergo radial excitations $n_\rho$ and $n_\lambda$.
Thus, the h.o. wave functions are characterized as $\phi_{l_\rho n_\rho; l_\lambda n_\lambda}$.

Resonances with $l_\rho + 2n_\rho + l_\lambda + 2n_\lambda = N$ belong to the $N^{\rm th}$ excitation shell.
The symmetry of the spatial wave function $\phi$ must ensure the overall symmetry of the combined space–spin–flavor wave function.

In the second excitation shell, four resonances with quantum numbers $I(J^P)=\nicefrac12(\nicefrac12^+)$ are expected.
The symmetry of their space–spin–flavor wave functions is given by
\begin{align}
^28[56]: &&(\phi_{S}\,\chi_{M_S}\,\tau_{M_S} + \phi_{S}\,\chi_{M_A}\,\tau_{M_A}); \nonumber\\  
^28[70]: &&(\phi_{M_A}\,\chi_{M_A}\,\tau_{M_S} - \phi_{M_S}\,\chi_{M_S}\,\tau_{M_S} \nonumber
        &+\phi_{M_S}\,\chi_{M_A}\,\tau_{M_A} - \phi_{M_A}\,\chi_{M_S}\,\tau_{M_A});\\
^48[70]: &&(\phi_{M_S}\,\chi_{S}\,\tau_{M_S} + \phi_{M_A}\,\chi_{S}\,\tau_{M_A}); \nonumber\\  
^28[20]: &&(\phi_{A}\,\chi_{M_A}\,\tau_{M_S} + \phi_{A}\,\chi_{M_S}\,\tau_{M_A})\,. \nonumber
\end{align}

The spatial wave function of $N^*$'s in the second excitation shell can be cast into the form
\begin{align}
^28[56]:&\quad \phi_{S}(l_\rho n_\rho; l_\lambda n_\lambda)&=& \ (\phi_{0100} + \phi_{0001})/\sqrt2\,;&\quad\nonumber\\
^28[70]:&\quad \phi_{M_S}(l_\rho n_\rho; l_\lambda n_\lambda)&=& \ (\phi_{0100} - \phi_{0001})/\sqrt2\,; & \phi_{M_A}(l_\rho n_\rho; l_\lambda n_\lambda) =  \phi_{1010}\,;   \nonumber\\  
^48[70]:&\quad \phi_{M_S}(l_\rho n_\rho; l_\lambda n_\lambda)&=& \  (\phi_{2000} - \phi_{0020}/\sqrt2)\,; & \phi_{M_A}(l_\rho n_\rho; l_\lambda n_\lambda) = \phi_{1010}\,;  \nonumber \\
^28[20]:&\quad  \phi_{A}(l_\rho n_\rho; l_\lambda n_\lambda)&=& \  \phi_{1010}\,.\qquad\ \quad\ & \label{20} 
\end{align}

\begin{table}[b!]
\footnotesize
    \caption{Masses and configuration mixing for the excited states of the nucleon with 
    $J^P=\nicefrac12^+$ in the second excitation shell from \cite{Isgur:1978wd} 
    (\cite{Loring:2001kx}). The masses are in MeV, the contributions in \%.\vspace{2mm}}
    \label{tab:mix}
\renewcommand{\arraystretch}{1.4}
    \centering\footnotesize
    \begin{tabular}{|ccccccc|}
    \hline
\hspace{-2mm}&\hspace{-2mm}M$_{\rm RPP}$\hspace{-2mm}&\hspace{-2mm}M$_{\rm calc}$\hspace{-2mm}&\hspace{-2mm}$^28[56]$\hspace{-2mm}&\hspace{-2mm}$^28[70]$\hspace{-2mm}&\hspace{-2mm}$^48[70]$\hspace{-2mm}&\hspace{-2mm}$^28[20]$\\\hline
$N(1440)$\hspace{-2mm}&\hspace{-2mm}1440\er30\hspace{-2mm}&\hspace{-2mm}1405 (1518) \hspace{-2mm}&\hspace{-2mm}\underline{97.0 (90.8)}\hspace{-2mm}&\hspace{-2mm}3.0 (6.1) \hspace{-2mm}&\hspace{-2mm} 0 (0.1)\hspace{-2mm}&\hspace{-2mm} 0 (0.2)\\
$N(1710)$\hspace{-2mm}&\hspace{-2mm}1710\er30\hspace{-2mm}&\hspace{-2mm}1705 (1729)\hspace{-2mm}&\hspace{-2mm}2.6 (15.5) \hspace{-2mm}&\hspace{-2mm}\underline{87.4 (81.4)} \hspace{-2mm}&\hspace{-2mm} 9.5 (0.7) \hspace{-2mm}&\hspace{-2mm} 0.5 (0.3) \\
$N(1880)$\hspace{-2mm}&\hspace{-2mm}1880\er50\hspace{-2mm}&\hspace{-2mm}1890 (1950)\hspace{-2mm}&\hspace{-2mm}0.4 (0.9) \hspace{-2mm}&\hspace{-2mm} 9.0 (0.9) \hspace{-2mm}&\hspace{-2mm}\underline{70.3 (88.6)} \hspace{-2mm}&\hspace{-2mm} 20.3 (7.4) \\
$N(2100)$&2100\er50&2055 (1996)& 0.0 (34.2) & 0.6 (5.9) & 20.2 (6.8) &\underline{79.2 (50.8)} \\
\hline
    \end{tabular}
\end{table}
These are harmonic-oscillator (h.o.) wave functions. The true wave functions, however, are mixtures of these basis states. In principle, the nucleon and states from the fourth excitation shell could also contribute to the wave function of a given resonance, but mixing between different shells is expected to be small and is therefore neglected.

The mixing parameters for the non-relativistic quark model of Isgur and Karl~\cite{Isgur:1978wd} and for the relativistic Bonn model~\cite{Loring:2001kx} are listed in Table~\ref{tab:mix}, together with the corresponding masses. Both models contain free parameters that are adjusted to reproduce the observed Breit--Wigner masses; hence, their predictions should be
compared to the experimental Breit--Wigner values.

Four states with quantum numbers $I(J^P)=\nicefrac12(\nicefrac12^+)$ are predicted, and indeed four such resonances are observed experimentally: $N(1440)\nicefrac12^+$, $N(1710)\nicefrac12^+$, $N(1880)\nicefrac12^+$, and $N(2100)\nicefrac12^+$. It is therefore tempting to identify these experimentally established resonances with the corresponding quark-model states.

\begin{figure}[!h]
\centering
\begin{tabular}{cc}
\hspace{-4mm}    \includegraphics[width=0.5\linewidth]{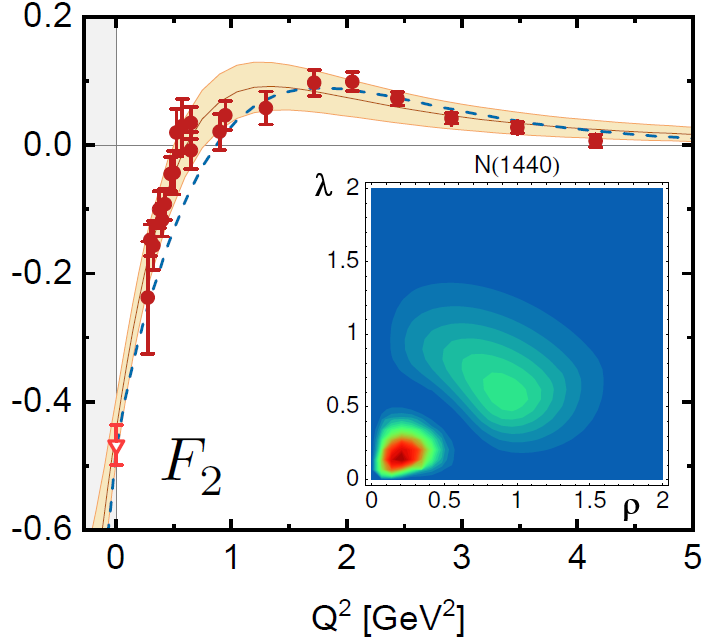}&
\hspace{-4mm}    \includegraphics[width=0.5\linewidth]{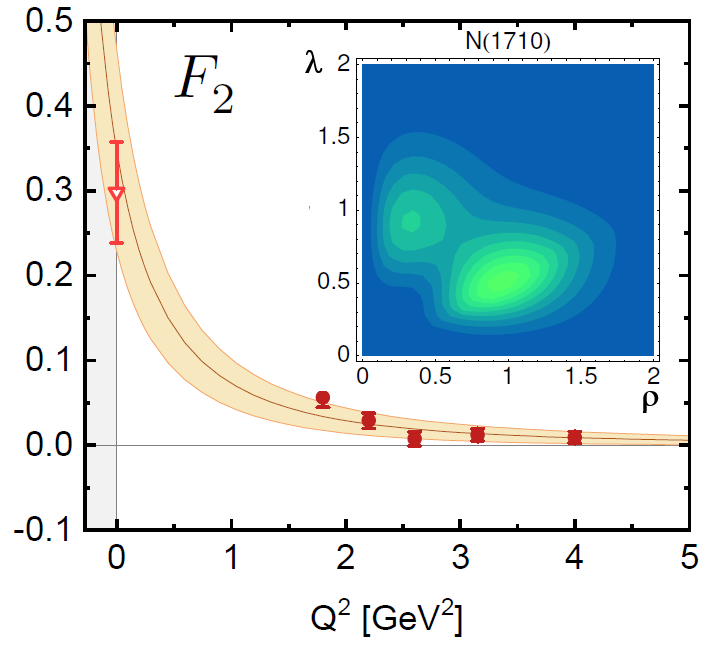}\vspace{-2mm}
 \end{tabular}
    \caption{\label{fig:2radials} Transition form factor $F_2(Q^2)$ for the $N(1440)\nicefrac12^+$  
    and $N(1710)\nicefrac12^+$.
    The data are from Refs.~\cite{CLAS:2012wxw,CLAS:2014fml}, the band is from~\cite{Eichmann:2018ytt} and the dashed curve
    from \cite{Drechsel:2007if}. The quark-model spatial probability distributions 
    from \cite{Melde:2008yr} are shown as insets.
    }
\end{figure}
Both models agree that the Roper resonance is predominantly the first radial excitation of the nucleon, with only small admixtures from higher configurations. Of course, the Roper may also possess a meson cloud, which is not accounted for in simple quark models. Nevertheless, its structure is largely governed by its three-quark core. This observation lends support to interpreting $N(1710)\nicefrac12^+$ as the second radial excitation~\cite{Brodsky:2014yha}. However, this conjecture is not supported by electroproduction data.

Figure~\ref{fig:2radials} compares the $N(1440)\nicefrac12^+$ and $N(1710)\nicefrac12^+$ transition form factors. The Roper transition form factor crosses zero due to the node in its wave function, with the position of the zero shifted by meson-cloud effects. For a second radial excitation, two zero crossings could be expected, but such behavior is not observed experimentally. In quark models, both $N(1440)\nicefrac12^+$ and $N(1710)\nicefrac12^+$ possess one node: the $N(1440)\nicefrac12^+$ in $\rho^2+\lambda^2$, and the $N(1710)\nicefrac12^+$ in $\rho^2-\lambda^2$ (see the inset in Fig.~\ref{fig:2radials}). The node
in $\rho^2-\lambda^2$ does not result in a zero crsossing of the transition amplitude.

The transition form factor for $N(1880)\nicefrac12^+$ is not yet known, but it appears plausible to assign this state to the expected member of a spin quartet with $L=2$ and $S=\nicefrac32$, coupling to $J=\nicefrac12$. This quartet should be accompanied by three additional states with $L=2$, $S=\nicefrac32$, coupling to $J=\nicefrac32$, $\nicefrac52$, and $\nicefrac72$. These may correspond to the experimentally observed resonances $N(1900)\nicefrac32^+$, $N(1860)\nicefrac52^+$, and $N(1990)\nicefrac72^+$.

The $N(2100)\nicefrac12^+$ resonance remains a puzzle. In a first interpretation, it could be assigned a dominant $^2[20]$ configuration. In the leading spatial component of this configuration, both oscillators are excited, see
$\phi_A$ in Eq.~(\ref{20}). In Ref.~\cite{Isgur:1978wd}, this contribution amounts to approximately 80\%, implying that production in a single-step process should be strongly suppressed. In contrast, Ref.~\cite{Loring:2001kx} finds a suppression by only about a factor of two.

Alternatively, $N(2100)\nicefrac12^+$ could be interpreted as a hybrid baryon.
Mesonic hybrids with an excited gluon field, often denoted as $q\bar qg$ mesons, were predicted~\cite{Horn:1977rq} 
soon after QCD was established as the theory of the strong interaction (see Ref.~\cite{Gross:2022hyw} for a review).
Mesonic hybrids may possess {\it exotic} quantum numbers—combinations that are not accessible to conventional $q\bar q$ systems.
They are predicted to decay primarily through string breaking, leading to two-meson final states in which one of the mesons carries intrinsic orbital angular momentum~\cite{Isgur:1985vy,Kokoski:1985is}.
This selection rule is broken when the quark and antiquark masses are unequal.
A candidate for such a state, the $\pi_1(1600)$ with $J^{PC}=1^{-+}$, has been identified unambiguously in its decays to $f_1(1285)\pi$~\cite{E852:2004gpn}, $b_1(1235)\pi$~\cite{E852:2004rfa,Baker:2003jh}, $\eta'\pi$~\cite{E852:2004gpn}, $\rho\pi$~\cite{COMPASS:2018uzl}, and $\eta\pi$~\cite{Kopf:2020yoa}; see Ref.~\cite{Grube:2022qcd} for a review.

\begin{wrapfigure}[25]{r}{0.5\textwidth}
    \centering
\includegraphics[width=\linewidth]{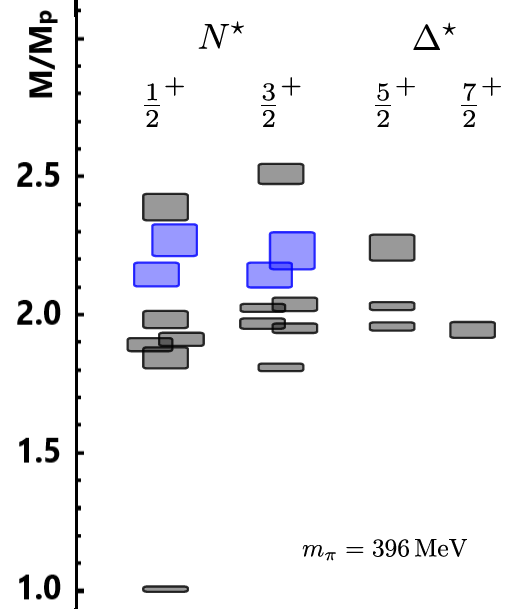}
\caption{(color online) Mass spectra of $N^*$ and $\Delta^*$ baryons. The mass scale is made to reproduce 
    the masses of the nucleon and of $\Delta(1950)\nicefrac72^+$. Baryons with a large hybrid content are
    highlighted in blue. Adapted from~\cite{Dudek:2012ag}.  }
    \label{Fig:lattice}
\end{wrapfigure}

Hybrid baryons, in contrast, do not possess exotic quantum numbers.
Their identification must therefore rely on comparisons with mass predictions and decay patterns.
Lattice QCD calculations have predicted the masses of hybrid baryons~\cite{Dudek:2012ag,Khan:2020ahz}.
Figure~\ref{Fig:lattice} shows the expected mass spectra of $N^*$ states with $J^P=\nicefrac12^+$ and $\nicefrac32^+$ and of $\Delta^*$ states with $J^P=\nicefrac52^+$ and $\nicefrac72^+$.
These calculations were performed with quark masses corresponding to a pion mass of about 400\,MeV; hence, all predicted masses are systematically too high.
In Fig.~\ref{Fig:lattice}, all masses are normalized to the nucleon mass.
The $\Delta(1950)\nicefrac72^+$ appears near 1900\,MeV, lending support to the calculations.

The four $I(J^P)=\nicefrac12(\nicefrac12^+)$ states discussed here are all observed below 2\,GeV.
Above this region, a mass gap is seen: the next $qqq$ state, belonging to the fourth excitation shell, is expected at about 2.4\,GeV.
Within this gap (2.0–2.4\,GeV), two hybrid baryons are predicted.
Thus, if the $N(2100)\nicefrac12^+$ is not a member of the [20]-plet, it could plausibly be interpreted as a hybrid baryon.
By analogy with hybrid mesons, one expects hybrid baryons to decay predominantly into a baryon–meson pair in which one of the final-state hadrons carries one unit of intrinsic orbital angular momentum.

A third possibility remains: $N(2100)\nicefrac12^+$ could represent the lowest-mass $J^P=\nicefrac12^+$ state in the fourth excitation shell, its mass lowered analogously to the Roper resonance in the second shell.
In this interpretation, it would belong to a $^28[56]$ configuration, and its wave function would be given by

\begin{equation}
\phi_{l_\rho n_\rho, l_\lambda, n_\lambda, } =
\frac{1}{\sqrt{6}}\,\phi_{02,00}  +\frac{1}{\sqrt{6}}\,\phi_{00,02} +\frac{\sqrt{5}}{3}\,\phi_{01,01} 
-\frac{1}{3}\,\phi_{20,20}\,.
\end{equation}
What do we expect?
For a member of the [20]-plet with a spatial wave function
$\phi_{l_\rho n_\rho,, l_\lambda n_\lambda} = \phi_{10,10}$,
we expect that a first de-excitation leads to spatial wave functions of the form
$\phi_{10,00}$ or $\phi_{00,10}$, corresponding to one unit of orbital angular momentum.
The two oscillators may also de-excite coherently through pion emission, where the two pions form a  $f_0(500)$.
Configuration mixing introduces components from $^28[56]$, $^28[70]$, and $^48[70]$,
which can decay into $N\pi$, $N\rho$, or $\Delta(1232)\pi$ final states without intrinsic orbital angular momentum.

For a hybrid state, we expect decays into a baryon–meson pair where one of the hadrons carries intrinsic orbital angular momentum.
Since the decay products have unequal masses, the symmetry that would otherwise enforce the angular-momentum distribution is broken, allowing also $S$-wave decays into channels such as $N\pi$, $N\rho$, or $\Delta(1232)\pi$.
Nevertheless, a substantial fraction of the decays should proceed through intermediate states such as $N(1535)\nicefrac12^-$, $N(1520)\nicefrac32^-$, or $Nf_0(500)$, where the latter represents correlated two-pion emission.

\begin{wraptable}[14]{l}{6.5cm}
\renewcommand{\arraystretch}{1.4}
{
\begin{tabular}{|c|c|c|c|} \hline
\hspace{-2mm}$N\pi$\hspace{-2mm}&\hspace{-2mm}$N\eta$\hspace{-2mm}&\hspace{-2mm}$N\eta'$\hspace{-2mm}&\\
\hspace{-2mm}16\er4\hspace{-2mm}&\hspace{-2mm}9\er4\hspace{-2mm}&\hspace{-2mm}4\er2   \hspace{-2mm}&\\\hline
\hspace{-2mm}$\Lambda K$\hspace{-2mm}&\hspace{-2mm}$\Sigma K$\hspace{-2mm}&\hspace{-2mm}$\Delta\pi$\hspace{-2mm}&\hspace{-1mm}$N\rho$\\
\hspace{-2mm}$<1$\hspace{-2mm}&\hspace{-2mm}7\er 3\hspace{-2mm}&\hspace{-2mm}10\er4\hspace{-2mm}&\hspace{-1mm}17\er7\\
\hline 
\hspace{-2mm} $N_{1535}\pi$\hspace{-2mm}&\hspace{-2mm} $N_{1710}\pi$\hspace{-2mm}&\hspace{-2mm} $N_{1720}\pi$\hspace{-1mm}&
\hspace{-2mm}$Nf_0(500)$\\
\hspace{-2mm}$<1$\hspace{-2mm}&\hspace{-2mm}10\er5 \hspace{-2mm}&\hspace{-2mm} 15\er5\hspace{-2mm}&\hspace{-1mm}28\er6\\
\hline
\end{tabular}
}
\renewcommand{\arraystretch}{1.0}
\caption{ Branching ratios (in \%) for decays of the $N(2100)\nicefrac12^+$~\cite{PMahlberg} (preliminary). 
The sum is (116\er13)\%.  }
\label{BR}
\end{wraptable}
If $N(2100)\nicefrac12^+$ is instead a Roper-like state belonging to the fourth excitation shell,
we expect contributions from components of the type
$\phi_{l_\rho n_\rho,, l_\lambda n_\lambda} = \phi_{20,20}$,
corresponding to intrinsic orbital angular momentum $L=2$ in one of the outgoing hadrons.
The $\phi_{01,01}$ component could contribute to decays into radially excited states such as
$N(1440)\nicefrac12^+$ or $N(1710)\nicefrac12^+$.
The $\phi_{20,20}$ component is likely to decay into
$N(1680)\nicefrac52^+,\pi$ in an $F$-wave,
$N(1720)\nicefrac32^+,\pi$ in a $P$-wave,
or $N,f_2(1270)$ in a $D$-wave.
The latter two components can also emit two pions coherently from the two excited oscillators,
leading to $Nf_0(500)$ as an intermediate state.

Table~\ref{BR} presents preliminary results from a BnGa coupled-channel analysis that includes new data from the CBELSA/TAPS experiment~\cite{PMahlberg}, both with and without polarization observables.
The decays listed in the first two rows may originate from the $\phi_{02,00}$ and $\phi_{00,02}$ components,
while the $Nf_0(500)$ decays can be interpreted as correlated two-pion emission from the
$\phi_{20,20}$ and $\phi_{01,01}$ configurations.
The $N(1720)\nicefrac32^+,\pi$ and $N(1710)\nicefrac12^+$ decay modes are naturally expected
for a Roper-like $J^P=\nicefrac12^+$ resonance belonging to the fourth excitation shell.

These interpretations are clearly suggestive.
There is no selection rule that would forbid the $\phi_{02,00}$ and $\phi_{00,02}$ components from decaying into orbitally or radially excited intermediate states,
and the $\phi_{20,20}$ and $\phi_{01,01}$ components could decay, for instance, into
$N(1720)\nicefrac32^+,\pi$, with subsequent rescattering of the $N(1720)\nicefrac12^+,\pi$ system into
$N\rho$ or $\Delta(1232)\pi$.
However, the most straightforward interpretation of the results is that
$N(2100)\nicefrac12^+$ represents a low-mass member of the fourth excitation shell.

In summary, the sequence of positive-parity $N^*$ resonances with $J^P=\nicefrac12^+$ — from the Roper resonance $N(1440)\nicefrac12^+$~~ through~~ $N(1710)\nicefrac12^+$, $N(1880)\nicefrac12^+$, and up to $N(2100)\nicefrac12^+$ — reveals a consistent pattern in the excitation spectrum of the nucleon.
The Roper resonance is now well understood as the first radial excitation of the nucleon, dominated by its three-quark core and dressed by a substantial meson cloud.
The higher states may naturally be interpreted as successive excitations within the same quark-model framework, with increasing contributions from orbital motion and configuration mixing.
While alternative interpretations, such as dynamically generated or hybrid baryons, cannot be excluded, the most coherent picture emerging from both quark-model and coupled-channel analyses identifies $N(2100)\nicefrac12^+$ as a low-mass member of the fourth excitation shell — a Roper-like state completing the observed pattern of nucleon excitations. 
\section*{Acknowledgements}
The work of V. Burkert was supported by the US Department of Energy under contract DE-AC-06OR23177.

\bibliography{Baryon}
\end{document}